# Physics and Logical openness in Cognitive Models

Ignazio Licata
*ISEM, Institute for Scientific Ethics and Methodology, Pa ,Italy*
ignazio.licata@ejtp.info

**Abstract**

It is here proposed an analysis of symbolic and sub-symbolic models for studying cognitive processes, centered on emergence and logical openness notions. The Theory of Logical Openness connects the Physics of system/environment relationships to the system informational structure. In this theory, cognitive models can be ordered according to a hierarchy of complexity depending on their logical openness degree, and their descriptive limits are correlated to Gödel-Turing Theorems on formal systems. The symbolic models with low logical openness describe cognition by means of semantics which fix the system/environment relationship (cognition *in vitro*), while the sub-symbolic ones with high logical openness tends to seize its evolutive dynamics (cognition *in vivo*). An observer is defined as a system with high logical openness. In conclusion, the characteristic processes of intrinsic emergence typical of "bio-logic" - emerging of new codes-require an alternative model to Turing-computation, the natural or bio-morphic computation, whose essential features we are going here to outline.

## 1.The Physics of Cognitive Processes

The Symbolic Approach to AI and the cognitivist models are well represented by the Newell and Simon Physical Symbolic System (Newell, 1990). It is essentially based on a "boxes and arrows" representation, diagrams providing a representation both of the system knowledge and its environment in terms of "information flow". Although providing precious suggestions to computational epistemology, such approach showed to be limited in a way that can be summarized by the peculiar kind of relationship linking cognitivism, "strong" IA and classical Physics. As everybody knows in Newton-Laplace Physics the energetic behaviors are fixed by Maxwell-Boltzmann Statistics, they are of two kinds:

Systems conserving information, characterized by conservation of energy which is expressed as the invariance of a Hamiltonian function according to a deterministic scheme;

Systems dissipating information, where the entropy increasing is detected, which is the sign of the growing distribution amplitude among the energetic levels and whose asymptotic "destiny" is predictable.

It is clear that such kind of Physics is unsatisfying to exhaustively describe mental processes. In particular, in cognitive models it is introduced an artificial difference between the "software" and the "hardware" level, which brings forward again the mind/body Cartesian dichotomy (Cartesian cut). The information elaborating is considered as a conservative process – deterministic and logically closed – and any dissipative feature is left to the "hardware" level supporting the symbolic one. We also notice the close connection between cognitivism and Turing-computability.

In this class of cognitive theories the information notion is "dispossessed" of its natural physical origin. On the contrary, for instance, cognitive maps and memory are not static representations, but variable resources of adaptive responses to stimuli. It suggests to take into consideration models where the intelligent system's informational analysis cannot leave out the intrinsic "dynamics" and "thermodynamics". So our attention is turned to a third class of systems:

Systems amplifying information, where dissipation and self-organization phenomena take place (Licata ,2003).

Without going here into mathematical details, it can be demonstrated that within this class new and unpredictable behaviors, started even by tiny fluctuations, can happen. Consequently it is within this class of models we have to search a description of cognitive processes which takes into account the dynamics and thermodynamics of computation. It is known that the non-linear neural nets are a particularly interesting class (Levine, 2000; Borghi-Iachini, 2002). Research on these models has pointed out that separating "mind" from its environment is impossible, so we are strongly tempted to propose the following definition for Intelligent System (IS):

*An Intelligent System is an information amplifying dynamic system associated to a computational body which allows the non-linear, parallel, distributed processing of information between system and environment.*

We introduce here a subtle but quite important difference between system and computational body, which will be useful to overcome the artificial opposition between cognitivism and connectionism in the frame of a unitary conception. To a careful theoretical examination, in point of fact, the real difference between these two approaches is not the greater "biological plausibility" of connectionism – which is real in a very limited sense – but rather the different description of the system-environment relationship. In other words, what characterizes the IS type is never a mere morphological and structural factor, but how such factor is functional in favoring an operational closeness relationship with environment (Maturana & Varela, 1998). In short, that's a kind of system's "permeability" to the various informational flow coming from environment, but able to maintain the system autonomy and the originality of adaptive responses. We can think an IS as a particular spatio-temporal distribution with computational ability showing autonomous behaviors within the wider informational field where it is immersed. Such system's capability is linked to the possibility of showing phase transition, self-organized criticality and emergence phenomena. There is a huge literature on phase transitions and self-organized criticality by now (see, for ex., Bushev, 1994), so we are focusing our attention on emergence.

## 2.Observers, Intelligent Systems and Emergence

It will be useful for our aims to choose here the systemic-cybernetic approach which focuses on relationship (system, environment) as fundamentally inseparable, except under opportune conditions suggested by the adopted experimental-observational framework. In other words, any physical description always implies an observer who operates the grouping/parting of the observed phenomena in systemic classes, endowed with global properties, which cannot be simply reduced to the constituent elements and susceptible of multiple levels of description. We can also say that an observer is a system whose dynamic inter-relation with environment defines new systemic classes. So we can consider the observer as an emergence detector. This is what an IS really does, therefore in order to build a general theory of cognition it is of paramount interest to dwell on the observer concept and the correlated notion of emergence.

We can define an emergent process, in general terms, as a dynamic process which modifies the correlations among the system' significant variables and operates a rearrangement of the representation and production rules of information inside the system. The consequence will be the appearing of new relations between environment and system in terms of input-output. Of course, it can happen at different levels of complexity.

Let us consider, for example, an autonomous model like the ones studied in evolutive robotics. They are computational devices led by a genetic algorithm and trained by a neural net which lets the robot explore the environment and optimize more and more its responses; they show up unpredictable behaviors simply on the basis of the system initial description. In this case it is clear that the emergent properties are the new faculties of the agent. If we give a complete description of the environment and a complete specification of the structure of the agent, it will be possible to get an algorithmic compression of the system (agent, environment) and any observed behavior can be seen as a *computational emergence* case completely specifiable within a syntactic information theory (Crutchfield, 1994; Heylighen, 1991; Baas-Emmeche,1997). What here appears as a "genuine" emergent property at a descriptive level, is formally describable by introducing a broader meta-level. There exists a more "radical" form of emergence which has been called *observational* (Baas-Emmeche, 1997; Cariani, 1991). With observational emergence the possibility to describe emergent properties by a simple computational model is excluded.

Let us imagine to insert an IS in a real environment describable in terms of ordinary physics and biology. The IS sensorial apparatus will detect an informational flow by means of an interaction which will modify irreversibly them both. If the IS is sufficiently complex, in order to use and elaborate information, the system has to convert it in multiple series of internal codes, which is to say it has to operate a *semantic appropriation of information.* It is here of great importance the notion of *code* as the system's ability to store and manage information in relation to its internal structure, its dynamic history and its relations with environment. Next, to be able to act on the external world, the *output* has to be converted again into the environmental code, with further dissipation of energy. When it happens, if we want to make forecasts of some sort on the IS behaviors, we cannot limit our investigation to its structural modifications, but we have to connect them to its world description, which is to say, we have to investigate its semantics.

It cannot be done in a relatively simple way as it was for the above mentioned robot. We can intuitively understand it if we consider that not only should it require a complete description of the world state but also of the IS structure' slightest detail and their co evolution modality.

Even supposing this new version of the *reductionist dream* as scientifically plausible, we have to keep in mind here that any physical description is centered on observer, and any measurement modify the couple observer/observed. So we should need an observer able to detect information without modifying it at all, practically void of internal structure, a *Laplacian demon* openly contrasting with the physics we know! This is what we mean when we state such kind of emergence is not algorithimically compressible and appears someway "irreducible", that's why it is also said *intrinsic emergence*.

At this point, naturally arises from the above-stated the definition:

*An Intelligent System (IS) is an observer able to detect systemic properties and to build dynamic world representations by means of intrinsic emergence processes associated with the existence of inner codes.*

The two proposed definitions, jointly considered, suggest some essential ingredients to build a general theory of intelligent systems up. We can synthesize them by introducing the logical openness and thermodynamic notions.

### 3. Emergence as logical openness and thermodynamics

Without any exhaustiveness or rigorousness pretence, we can identify the "capability" of an IS as the richness of its emergent processes connected to the interrelation with environment.

By studying the living systems, we know that the thermodynamic openness, which is the IS capability of being "permeable" to the matter/energy flow, is a necessary but not sufficient condition. The dissipative structures and the self-organized criticality processes are classical examples by now (Bak, 1996). The mathematical description of such systems can be carried out without much difficulty, and there exist a series of very general results which set strict limits to the complexity of the structures we can get by following this criterion. In particular, for our purposes, it is useful to underline that in dissipative systems the emergence of structures can be completely defined in terms of computational emergence, since this kind of system totally falls under a *logical closed model* such that:

there is a whole description of the relations among the state variables;

it is explicitly - and precisely as we desire - possible to define the interactions between system and environment.

Such characteristics allow fairly accurate predictions on the system evolution and its structural features, so they don't imply the existence of intrinsic emergence processes. Those ones need a further condition, *logical openness,* which can be described by introducing a hierarchy of the possible system/environment relations (Minati, Penna, Pessa, 1998; Licata, 2003). A formally complete theory of this kind of systems is still a far goal, but we can here outline some essential features.

Let us consider, for example, the case of system/environment interactions depending on the system's *internal* state; it can happen in the form of *values*, such as in the case of phase transitions, as well as when the *form* itself of these interactions depends on the system's responses. In the first case, we speak of system with level one logical openness, in the second of level two. The latter level can be considered as the indicator of the system's ability to unpredictably manage information by actively operating on the external world in a way not merely ascribable to the initial model. This is a typical behavior of intrinsic emergence phenomena.

We can put it like *the system ability of playing a different play from what the model planned*; therefore it is related to its capability to express new semantic domains. So we can identify the intrinsic emergence with the system aptitude to produce knowledge representations. As for logical open systems, our being precluded from getting a whole description of the system's internal states and the consequent irreversible modification of the observed behavior are key points. They naturally lead us to take into consideration a series of *indetermination principles* intrinsically connected to the study of cognitive and biological processes(Vol'kenshteĭn,1988). Thus in a General Theory of IS we can expect the occurring of indetermination relations, the higher the system's logical openness the more constraining they will be.

Generally, we define a system endowed with *n level logical openness* if it can be characterized by *at least* a n number of constraints, with finite n. Thermodynamically speaking, such definition finds its immediate significance if we consider that the more structured the system the more the maintenance of this structure has to tackle the dissipation required by the thermodynamic openness. The system maintains its autonomy thanks to a series of constraints; we can intuitively understand that the number of constraints shows the thermodynamic compromise between system and environment. Such constraints are generally functions like $f(x_n,t)$, where the values $x$ and $f(x,t)$ will be variably distributed over time between system and environment. To be more precise, domain and co-domain of these functions are not elements of a system-defined set, but are defined on a union of sets including all the possible partitions of system/environment relations, so as to ideally keep into account the whole inter-relations between the two poles of the systemic description.

It can be shown that (a) a logical open system admits several complementary formal descriptions, and (b) each description of a logical open system by means of a model of n-fixed logical openness, with n completely specified constraints, gets a limited domain of validity, i.e. it can grasp only a little part of the system/environment informational processes. Globally considered both (a) and (b) are the systemic corresponding of Gödel and Turing incompleteness famous theorems. In particular, (a) justifies the indeterminacy relations, which are to be considered as indications for the optimal model choice for the intended aim.

After delineating the essential concepts of logical openness theory, we can evaluate in unitary way the cognitive and connectionist models, pointing out their limits and qualities, and overcome an old and artificial antithesis. The cognitivist models are very effectual when we deal with a low level of logical openness, with a n number of constraints depending neither on time nor on system state; in classical IA it is the traditional notion of "micro-world". On the other hand, the connectionist ones are more useful to grasp some essential features of the system emerging complexifying related to the system/environment dynamics when a greater degree of logical openness occurs. Only when a cognitive process emerges, is detected and fixed over time within a defined context can we describe it in symbolic terms. In a sense, the very cognitivism usefulness lies in its being applicable *far from the emergence zones*; on the contrary, the neural and sub-symbolic approaches are useful just during the dissipation-amplification phases of information, when the intrinsic emergence processes give rise to new codes, that is to say to new way to manage information. By using a biological metaphor, we could say that symbolic systems study cognition *in vitro*, whereas connectionism provides tools to understand cognitive processes *in vivo*. Similarly, we can compare cognitivism to thermodynamics at equilibrium, while connectionism corresponds – not metaphorically – to the study of processes far from equilibrium.

There exist many theories proposing criteria of comparison and compatibility between what – in the light of the above-stated - appear to be nothing but two different descriptive levels of cognitive acts (Smolensky, 1987; 1992; Clark, 1989). We only point out here that the logical openness theory requires a theoretical frame able to treat system and environment as a whole, rigorously defining the intrinsic emergence processes connected to the informational and energetic configuration and providing a detailed formal definition of IS as observer able to operate systemic choices. The most interesting researches in this direction have been provided by the tools of Quantum Field Theory (Ricciardi & Umezawa, 1967; Pessa & Vitiello, 1999).

## 4. Codes, Natural Computation and Turing-Machines

The relevant feature of both cognitive processes (Licata, 2003) and biological organisms (Barbieri, 2003) can be synthesized by saying that in such systems the interrelations with environment give rise to intrinsic emergence phenomena displayed as the system's capability to manage information in a new way, so widening its semantic domain. It corresponds to the appearing of new codes; we are not going to formally define them here, but we can think them as constraints imposed on the system informational flow and able to rearrange it in order to achieve a goal. It naturally lead us to query whether the Turing-Computability actually is the most suitable to grasp the peculiar features of such processes, or whether we should regard it as a particular case within a more extensive computation theory linked with the observer's choice and the kind of logical openness considered.

In its broadest sense computation is a mathematical relation associating a set of inputs with a set of outputs. The choice that is made on the basis of both the mathematical relation and the formats of input and output defines a *computational model* to analyze information. It is worthy noticing that any physical system exchanges information with environment, and it is defined by the peculiar way it does that; so it falls within the three categories described on § 1. Moreover, information can be directly linked to the fundamental concept of distribution of energetic levels. Any physical system is an information processor; such thing is a much more general notion than the computation one, which only pertains the way we choose to manage the exchanged information.

It is universally known that Alan Turing developed his computation concept so as to define in "operative" way the notion of actually computable function; he shrewdly analyzed what a mathematician (an IS!) does when working on such task. Later, following the spirit of Turing work, it has been pointed out how the Turing Machine (TM) notion is absolutely compatible with the known laws of physics (Minsky, 1972; Gandy, 1980; Davis, 1982; Arbib, 1987). Which thing does not imply that it is the only plausible computation model as it has all too often been wrongly taken for granted. Suffice here to remember that the T-comp. is an essentially discrete and countable process, syntactically defined, but undefined as to the spatio-temporal features of processing. On the contrary, in biological and cognitive processes the space-time modalities of computation are essential. This is a particular important aspect in soft-computing tools (Kosko, 1992; Zadeh, 1998), which can theoretically be reduced to Turing-comp.-based analysis – neural nets, cellular automata, genetic algorithms, fuzzy logic, etc. - , but they show, in fact, extremely different "vocations" in analyzing the informational flow. The equivalence between these operational tools and TMs is carried out by analyzing *a posteriori* the work done, without taking into consideration the specific distribution of the computational activity. It immediately leads us back to the distinction between systems with different logical openness. We noted there that symbolic models represent knowledge as once and for all fixed in space and time, while the connectionist approach showed to be more useful when we had to focus on emergent processes where the spatio-temporal dynamic plays a decisive role. Similarly, *it is because of its very general definition that Turing-computation is apter to analyze an already codified informational flow!* On the contrary, the cognitive processes are the place *par excellence* for adaptive strategies, phase transitions and intrinsic emergence; that is to say they are the best suited for the emerging of new codes "mirroring" the system/environment dialogue in its dynamic evolving instant by instant. Moreover, physical processes can always been studied from manifold viewpoints, and there is no reason, either theoretical or experimental, to prefer a discrete framework to a continuous one and viceversa. On the other hand, we can consider neither of them as approximations, given the huge difference between the continuous and discrete mathematical approaches. A TM is a discrete and countable tool, while a neural net is described by a system of differential equations based on continuous functions. So we should ask whether a discrete computational model is useful when we deal with so powerful a tool as the continuum. We cannot fail to mention here some interesting cases related to quantum and classical physics which hinted that non-computability in Turing sense had not to be regarded as a theoretical "checkmate", but as an invitation to adopt a different theoretical approach.

Such considerations lead us to enunciate a new proposition:

*An Intelligent System (IS) is an observer able to choose different computational models for managing information in relation to its goals.*

A worthy consequence of the above statement is:

*An observer is a system with high logical openness.*

In so doing theory provides a natural context for the notions of "observer" and "observer's choices".

A rapidly developing research field is the study of the computational features of *continuous formal systems* which finds its ideal application within the ambit of neural nets and dynamic systems (MacLennan, 1992; Siegelmann,1999). It is called analog or natural computation because of its mathematical features and application fields in opposition to the discrete and countable features of TMs. The general characteristics of these approaches are related to the study of continuous computational fields, or the ones whose density of elements makes them suitable to be treated as continuous. Such kind of field can be formally defined by means of an Hilbert space, well known in Quantum Mechanics. The key points for a critical comparison with Turing comp. are:

*The involved quantities are continuous*. Both the cases where differentiation is never lacking and the "pathological" ones are taken into consideration, moreover it is always possible a discretization procedure so to get a not-continuous scenario;

*Information is treated by images,* that is to say continuous patterns distributed over space and time, so to treat what in a natural system corresponds to continuous gradation and variation by nuances;

*Noise and uncertainty are always present*. In a TM the imprecision of a very single symbol can compromise the whole computation; in natural computation, instead, such elements are a *resource* which includes all the dissipative processes involved both in the acquisition of information from environment by sensors and in the performing of a strategy by effectors.

*The computation process is endless and adaptive*. Physical and biological systems process information not only in relation to a formally defined and fixed task but continuously. The end of a computational activity has a different meaning in TM than it has in physical systems (problem solving and end of system by dissipation). The natural computation systems do not solve problems in formal sense, but endlessly elaborate responsive and adaptative strategies;

*Images represent nothing*. This is a key point and an important consequence of the theory. Differently from TMs, the images which are processed in a natural computation system are not symbols, but they correspond to the agent behavior and its goals, consequently their interpretation is not a fixed one. Any kind of interpretation varies over time, so we get a gradation of meanings versus the rule-fixed meanings typical of Turing comp.

*Codes and meaning domains are immanent to system,* i.e. instant by instant they mirror its structure and its interrelations with environment.

Many researches have already provided fruitful indications about the computational modalities of continuous formal systems, especially for their efficiency when compared to TMs as well as their ability to solve problems which are undecidible within Turing comp. (*halting problem*). We can intuitively realize that such capabilities have a strong connection with the system topology (interconnection, non-linearity, etc.) and its logical openness degree. Sub-Turing and Super-Turing results, globally considered, show that these systems are qualitatively different from TMs as for computational modalities and efficiency. Which thing can contribute to lead the debate on the singularity of human cognition within the scientific sphere, so avoiding any banal anthropocentrism.

## 5. Conclusions

A General Theory of IS will have to overcome the artificial contraposition between cognitivism and connectionism by formally developing a theory of observers endowed with logical openness and able, by means of intrinsic emergence processes, to produce new codes which lead the system in building world representations centered on its goal. A new concept of computation, and the consequential exploring of suitable computational tools, will be crucial in this kind of theory. More than a reason leads us to think that a really efficacious formalism to delineate a theory of IS could be a fit "semantic" widening of quantum theory (Dissipative Quantum Model of Brain, see for ex. Vitiello,2001).

It could be the formal core of a new *Physis* able to comprehend mind and matter as dynamical elements of a unitary scenario